\documentstyle[12pt,epsf]{article}
\renewcommand{\bar}[1]{\overline{#1}}

\begin{document}

\begin{flushright}
GEF-Th-2/2000\\
\end{flushright}

\bigskip\bigskip
\begin{center}
{\large \bf Transverse Spin Asymmetries in Drell-Yan Processes:\\
  $\mathbf p^{\uparrow} p \rightarrow \mu^+ \mu^- X$}
\end{center}
\vspace{12pt}

\begin{center}
 {\bf Elvio Di Salvo\\}

 {Dipartimento di Fisica and I.N.F.N. - Sez. Genova, Via Dodecaneso, 33 \\-
 16146 Genova, Italy\\}
\end{center}

\vspace{10pt}
\begin{center} {\large \bf Abstract}

\end{center}

We consider two different spin asymmetries in Drell-Yan processes generated by 
the collisions of an unpolarized proton beam on transversely polarized 
protons: the muon helicity asymmetry and the left-right asymmetry. We calculate 
the asymmetries in the framework of the QCD improved parton model, taking into
account the parton transverse momentum and considering the first order QCD 
corrections. The muon helicity asymmetry is sensitive to the quark transversity 
distribution and is nonvanishing even at zero order. On the contrary the
left-right  asymmetry vanishes at zero order but not at first order in the QCD 
coupling constant, as a result of gluon contribution.

\vspace{10pt}

\centerline{PACS numbers: 13.85.Qk, 13.88.+e}

\newpage

\section{Introduction}

$~~~~$ A major problem concerning the spin structure of the proton is the 
possibility of inferring the quark transversity distribution[1-5]
from  data[6, 7]. As pointed out by various authors[8-12], it is
quite  difficult to realize an experiment that is definitely sensitive to this 
quantity, which we shall denote by $h_1^f (x)$, $f$ being the quark flavors. 
Indeed

$~~i)$ $e - p$ deep inelastic scattering yields terms which are proportional to
$h_1^f$ but  suppressed like $m_f/Q$, where $m_f$ is the current quark mass 
and $Q$ the mass of the virtual photon.

$~ii)$ The double transverse spin asymmetry\cite{cpr,ma} in Drell-Yan 
processes is proportional  to the product $h_1^f (x) 
{\overline h}_1^f (x')$\cite{ba}, which is probably  very small, since we 
believe that, similarly to the helicity distributions, 
$|{\overline h}_1^f (x)| << |h_1^f (x)|$.

$iii)$ Other suggested\cite{ja1,ka} or planned\cite{ko} experiments are rather 
complicated and may present serious mishaps.

Up to now we can make only a rough evaluation of $h_1^f$\cite{ef}, based on 
the recent results of seminclusive $\pi$ electroproduction from HERMES 
and SMC\cite{her,smc}. Therefore doubts have been cast on the possibility of 
determining $h_1^f$\cite{ja2}. In the present situation, the best we 
can do is to search
for  all possible spin asymmetries, which are related to the transversity 
distribution, and to try to extract as much as possible information from the
data, when it will be available. 
The aim of the present paper is to suggest a new experiment 
sensitive to $h_1^f$. We propose to observe a muon pair produced by a collision 
between a transversely polarized proton and an unpolarized one, detecting, by 
means of standard experimental techniques\cite{bol}, the muon helicity 
asymmetry $A_1$ - that is the average longitudinal
polarization - of at least one of the muons (say $\mu^-$). This kind of 
proposal is complementary to a typical double polarization experiment ({\it e.
g.} inclusive DIS with polarized proton target and electron beam) and is
somewhat analogous to Collins' idea\cite{coll}, which consists in detecting the 
final quark polarization through the azimuthal asymmetry of the fragmentation 
function.

Spin asymmetries (especially with a single transverse spin)[19-33] are being  
currently considered in the literature\cite{ztb}, in view of experiments 
realized at FNAL\cite{fna} and DESY\cite{her}, or planned 
at RHIC[37-40] and other facilities\cite{ma,ber}. The aim of such 
proposals is to obtain the magnitude of twist-three terms, from which the spin 
asymmetries are expected to generate a nonzero contribution. The calculation
of  such terms involves technical difficulties and not all authors agree on
the  final results\cite{bhj}. Of particular interest is the left-right 
asymmetry\cite{bo1,ha}, often called single transverse spin asymmetry, which
can be determined from the  Drell-Yan collision described above by detecting
only the muon momenta. This  asymmetry - which we call $A_2$ - is nonzero only
if we consider, at least,  the first order correction in the QCD coupling
constant $g$, for which the contribution of a soft gluon pole plays a major
role. However  two different calculations\cite{bo1,ha} do not lead to the same 
formula.

We calculate asymmetries $A_1$ and $A_2$ at tree level, assuming the QCD parton 
model\cite{si} (see also\cite{an1}), and successively inserting "real" gluon 
corrections, for which we adopt an axial gauge\cite{et,ddt}, allowing us to 
develop a formalism as close as possible to the parton model. In calculating
the gluon corrections, we introduce the quark-quark correlation functions,
typical nonperturbative quantities. Our approach is different from the one
by other authors\cite{qi1,bo1,ha}, who adopt the formalism of quantum field
theory. However we match the two ways, obtaining a precise
definition of the correlation function in terms of destruction and creation
operators. This suggests a general procedure for calculating higher twist
contributions in parton model.

Our results on $A_2$ substantially agree with those of ref.\cite{bo1}, however 
the {\it gluonic pole} contribution is unfounded. In our 
approach the correlation functions are 
interpreted in terms of transverse quark and gluon polarization; moreover they 
can be evaluated quantitatively within a model by Qiu and Sterman\cite{qi1},
which we extend in quite a natural way to the case of polarized correlation 
functions.
The muon helicity asymmetry is strictly related to the transverse momentum 
dependent transversity distribution, whose integral over transverse momentum is 
$h_1^f$. Furthermore we observe that neglecting transverse momentum implies the
vanishing of the average helicity of the final muons, as a consequence of 
parity conservation. In this connection we like to stress the importance of the 
transverse momentum of the partons in various asymmetries\cite{an1}.

The Drell-Yan cross section is generally complicated [42-44]; it 
becomes considerably simpler\cite{co2,co3} - and in particular it may be 
approximated by a convolution over transverse momentum\cite{efp,mos} - if we 
limit ourselves to relatively large transverse momenta of the muon 
pairs\cite{mos}. This limitation is not a serious obstacle for our aims and it 
is even particularly appropriate for determining $h_1^f$, as we shall see.

In sect. 2 we define the two asymmetries we want to calculate and we give 
some general formulae for the asymmetries, the Drell-Yan cross section and the 
expression of the 
leptonic tensor. Sect. 3 is dedicated to the calculation of the hadronic 
tensor in the QCD parton model. In sect. 4 we write the QCD first order 
corrections by adopting an axial gauge. Moreover we impose gauge 
invariance and establish some symmetry properties for the correlation 
functions. Then we illustrate the properties of the perturbatively
calculable ("hard") coefficients; lastly we perform 
the calculations. In sect. 5 we give 
the expressions of the asymmetries and evaluate their orders of magnitude. 
In sect. 6 we draw some short conclusions.

\section{General Formulae}

$~~~~$ In Drell-Yan processes generated by the collisions between an
unpolarized  proton beam and a transversely polarized one, $p^{\uparrow}p \to
\mu^+ \mu^-  X$, two different kinds of asymmetries can be defined, {\it
i.e.}, the
muon helicity asymmetry, if the polarization of one final muon is detected,
and the  left-right asymmetry, if only final momenta of the muons are 
determined.

$~i)$ The muon helicity asymmetry is defined as

\begin{equation}
A_1 = {{d\sigma_+ - d\sigma_-} \over {d\sigma_+ + d\sigma_-}},
\label{as1}
\end{equation}
where ${d\sigma_{\pm}}$ is the inclusive Drell-Yan differential cross section 
with the final $\mu^-$ having positive (negative) helicity.

$ii)$ The left-right asymmetry is 

\begin{equation}
A_2 = {{d\sigma_r - d\sigma_l} \over {d\sigma_r + d\sigma_l}},
\label{as2}
\end{equation}
where ${d\sigma_{r(l)}}$ is the differential cross section with a final $\mu^-$ 
at the right (left) of the plane determined by the momentum and by the spin of
the polarized proton. Indeed, if the muon helicities are not detected, the only 
way of constructing an asymmetric term ({\it i.e.}, containing the $\varepsilon$ 
tensor) is  

\begin{equation}
A_2 \propto {\overline S} \cdot p,
\label{asy}
\end{equation}
where $p$ is the $\mu^-$ four-momentum and

\begin{equation}
{\overline S}_{\alpha} = \varepsilon_{\alpha \beta \gamma \delta} n_1^{\beta}
n_2^{\gamma} S^{\delta},
\label{sbar}
\end{equation}
$S$ being the Pauli-Lubanski four-vector of the polarized proton. Here we also 
have set

\begin{equation}
n_{1(2)} \simeq {1 \over \sqrt{2}} {P_{1(2)} \over |{\bf P}_{1(2)}|} 
\label{verso}
\end{equation}
and $P_1$ and $P_2$ are the four-momenta of the two protons, whose spatial 
parts are ${\bf P}_1$ and ${\bf P}_2$. Proton 1 is polarized, whereas 2 is not.
In the laboratory frame, where ${\bf P}_1$ =
-${\bf P}_2$, eq. (\ref{asy}) implies that the number of $\mu^-$ falling at the
right of the  plane determined by the spin and momentum of proton 1 is
different from those  which occur at the left of that plane.

The Drell-Yan cross section reads
 
\begin{equation}
d\sigma = {1 \over {4 P \sqrt{s}}} {{e^4} \over {Q^4}} L_{\mu\nu} H^{\mu\nu} 
\ d\Gamma,
\label{dsg}
\end{equation}
where $d\Gamma$ is the phase space element, whose expression is

\begin{equation}
d\Gamma = \frac{1}{(2\pi)^2} d^4 p ~ \delta(p^2) ~ \theta(p_0) ~~ 
d^4 {\overline p} ~ \delta({\overline p}^2) ~ \theta({\overline p}_0) ~~ 
\delta^4(p+{\overline p}-q).
\label{mps}
\end{equation}
Here $p$ and ${\overline p}$ are the four-momenta of $\mu^+$ and $\mu^-$ 
respectively, $q$ the four-momentum of the virtual photon, $Q$ its effective mass, $P$ = 
$|{\bf P}_1|$ = $|{\bf P}_2|$ the modulus of the momentum of each proton in the 
laboratory frame, $s$ the overall energy square, $L_{\mu\nu}$ the leptonic tensor 
and $H_{\mu\nu}$ the hadronic tensor.

The leptonic tensor is, in the massless approximation, 

\begin{equation}
L_{\mu\nu} = \frac{1}{4} Tr[\rlap/p(1\pm\gamma_5)\gamma_{\mu}\rlap/{\bar p}
\ \gamma_{\nu}],
\label{lept}
\end{equation}
$\pm$ being the sign of the the $\mu^-$ helicity. Then we can write

\begin{equation}
L_{\mu\nu} = S^l_{\mu\nu} \pm i A^l_{\mu\nu},
\label{lept1}
\end{equation}
where $S^l_{\mu\nu}$ and $i A^l_{\mu\nu}$ are, respectively, the symmetric 
(real) part and the antisymmetric (imaginary) part of the leptonic tensor:

\begin{eqnarray}
S^l_{\mu\nu} &=& p_{\mu} {\overline p}_{\nu} + {\overline p}_{\mu} p_{\nu} - 
g_{\mu\nu} p \cdot {\overline p}, \\
A^l_{\mu\nu} &=& \varepsilon_{\alpha\mu\beta\nu} p^{\alpha} 
{\overline p}^{\beta}.
\end{eqnarray}
Concerning the hadronic tensor, we set

\begin{equation}
H_{\mu\nu} = H_{\mu\nu}^{(0)} + H'_{\mu\nu},
\label{hhtt}
\end{equation}
where $H_{\mu\nu}^{(0)}$ and $H'_{\mu\nu}$ are, respectively, the QCD zero order 
and first order approximations, which will be calculated in the next two 
sections.

\section{Drell-Yan Hadronic Tensor in the QCD \\
Zero Order Parton Model}

Now we write the hadronic tensor for the high energy Drell-Yan process. 
According to the QCD parton model\cite{si}, we take into account the transverse 
momentum of the quark inside the hadrons. Moreover, as already explained in the 
introduction, we limit ourselves to time-like photons with transverse momenta 
$|{\bf q}_{\perp}|$ of order 1 $GeV$ and with $Q >> |{\bf q}_{\perp}|$, where 
$Q^2 = q^2$ and $q$ is the four-momentum of the pair. Then the generalized 
factorization theorem\cite{qi1}, in the covariant formalism\cite{land}, yields

\begin{equation}
H^{(0)}_{\mu\nu} = \sum_{f=1}^3 e_f^2 H_{\mu\nu}^f , 
\label{ht0}
\end{equation}
where $f$ are the three light flavors ($u,d,s$), $e_1$ = 2/3, $e_2$ = $e_3$ = 
-1/3, and 

\begin{equation}
H_{\mu\nu}^{f} = \int d\Gamma_q \sum_{T_1, T_2} 
[ q^{f}_{T_1} (p_1) 
{\overline q}^{f,2}_{T_2} 
(p_2) h^{\tilde{T}_{12}}_{\mu\nu} 
(p_1, p_2; S) + (1 \leftrightarrow 2)]. 
\label{hadt}
\end{equation}
Here $q^{f}_{T_l}$ (${\overline q}^{f}_{T_l}$) ($l$ = 1,2) are the 
probability density functions of finding a quark (antiquark) in a pure spin 
state whose third component along the proton spin is $T_l$. Moreover 
$d\Gamma_q$ 
is the analog of the element of phase space of muons (see eq. (\ref{mps})):

\begin{equation}
d\Gamma_q = \frac{1}{(2\pi)^2} d^4 p_1 ~ \delta(p_1^2) \theta(p_{10}) ~~ 
d^4 p_2 ~ \delta(p_2^2) ~ \theta(p_{20}) ~~ \delta^4(p_1+p_2-q)
\label{qps}
\end{equation}
and $p_l$ are the four-momenta of the two active partons. These are taken on 
shell and massless, which is a good approximation for the values of
$|{\bf q}_{\perp}|$ considered. Lastly

\begin{equation}
h^{\tilde{T}_{12}}_{\mu\nu} = \frac{1}{3} Tr (\rho^{T_1} \gamma_{\mu}
{\overline \rho}^{T_2} \gamma_{\nu}), 
\label{form1}
\end{equation}
where the factor 1/3 comes from color averaging in $q$-${\bar q}$ 
annihilation, $\tilde{T}_{12} \equiv (T_1, T_2)$ 
and $\rho$ is the spin density matrix\cite{la}, 

\begin{equation}
\rho^{T_l} ({\overline \rho}^{T_l}) = {1 \over 2} \rlap/p_l [1 + 2 T_l 
\gamma_5 (\pm \eta_{\parallel} + \rlap/\eta_{\perp})].
\end{equation}
$2T_l\eta_{\parallel}$ and $2T_l\eta_{\perp}$ are respectively the helicity and 
the transverse Pauli-Lubanski four-vector of the active partons. We have
$\eta_{\parallel} = {\bf S} \cdot {\bf n}_l$, where ${\bf n}_l = {\bf p}_l / 
|{\bf p}_l|$ and ${\bf S}$ and ${\bf p}_l$ are, respectively, the space 
components of $S$  and of $p_l$ in the laboratory frame. Moreover 
$\eta_{\perp} \equiv (0, {\bf S} - 
\eta_{\parallel} \ {\bf n}_l)$ ($l = 1,2$) and  $\eta_{\parallel}$, a Lorentz 
scalar in the limit of zero quark mass, can be defined covariantly \cite{ar2}.

Carrying on the integration (\ref{hadt}) over the time and longitudinal 
components of $p_1$, and adopting the light cone formalism, we get, in the 
limit of high $Q^2$ and $|{\bf p}_{l\perp}|$ $<<$ $Q$ ($l$ = 1, 2),

\begin{equation}
H_{\mu\nu}^{f} = \frac{1}{4\pi^2 Q^2}\int d^2 p_{1\perp} \sum_{T_1, T_2} 
[ q^{f}_{T_1} (x_1, {\bf p}_{1\perp}) 
{\overline q}^{f}_{T_2} 
(x_2, {\bf p}_{2\perp}) h^{\tilde{T}_{12}}_{\mu\nu} 
(x_1, x_2; S) + (1 \leftrightarrow 2)].
\label{hadt1}
\end{equation}
Here $x_{1,2}$ = $(q_0 \pm q_{\parallel})/{\sqrt{s}}$ are the longitudinal 
fractional momenta of the two active partons, $q_0$ and 
$q_{\parallel}$ are respectively the time and longitudinal component of $q$, and
\begin{equation}
{\bf p}_{2\perp} = {\bf q}_{\perp} - {\bf p}_{1\perp}. 
\end{equation}

Since the proton 2 is unpolarized, we set

\begin{equation}
q^{f}_{(T_2 = 1/2)} = q^{f}_{(T_2 = -1/2)}, ~~~~~~ ~~~~~~ ~~~~~~~
{\overline q}^{f}_{(T_2 = 1/2)} = {\overline q}^{f}_{(T_2 = -1/2)}.
\label{rel2}
\end{equation}
Inserting eqs. (\ref{form1}) to (\ref{rel2}) into eq. (\ref{hadt}), and taking 
into account the relation 

\begin{equation}
Q^2 = 4 x_1 x_2 P^2,
\label{krel}
\end{equation}
we get

\begin{equation}
H_{\mu\nu}^f = S_{\mu\nu}^f + i A_{\mu\nu}^f, 
\label{ht2}
\end{equation}
where $S_{\mu\nu}^f$ and $i A_{\mu\nu}^f$ are respectively the symmetric and 
antisymmetric part of the hadronic tensor:

\begin{eqnarray}
S_{\mu\nu}^f &=& \frac{1}{24\pi^2} s_{\mu\nu} Q^f, 
\label{symte}
\\
A_{\mu\nu}^f &=& \frac{2\sqrt{x_1x_2}}{24\pi^2Q} a_{\mu\nu} \delta Q^f, 
\label{asmte}
\end{eqnarray}
where
\begin{eqnarray}
Q^f &=& \int d^2 p_{1\perp} [q^f_1(x_1, {\bf p}_{1\perp}^2) 
{\overline q}^f_2 (x_2, {\bf p}^2_{2\perp})+ (1\leftrightarrow 2)], 
\label{qf}
\\
\delta Q^f &=& \int d^2 p_{1\perp} 
\frac{{\bf S}\cdot {\bf p}_{1\perp}}{x_1} 
[\delta q^f_1 (x_1, {\bf p}_{1\perp}) {\overline q}^f_2 
(x_2, {\bf p}^2_{2\perp}) - \delta {\overline q}^f_1
(x_1, {\bf p}_{1\perp}) q^f_2 (x_2, {\bf p}^2_{2\perp})].
\label{dqf}
\end{eqnarray}
Here we have introduced the distribution functions:

\begin{equation}
q^f_l  = \sum_{T_l = -1/2}^{1/2} q^{f,l}_{T_l}, \ ~~~~ \ ~~
\delta q^f_l = \sum_{T_l = -1/2}^{1/2} 2T_l q^{f,l}_{T_l},
\end{equation}
and similarly for the antiquarks. Furthermore

\begin{equation}
s_{\mu\nu} = n_{1\mu} n_{2\nu} + n_{2\mu} n_{1\nu} - g_{\mu\nu}, ~~~~~~~~~
a_{\mu\nu} = \varepsilon_{\alpha\mu\beta\nu} n_1^{\alpha} n_2^{\beta}.
\end{equation}
Such tensors fulfil gauge invariance up to twist-4 terms.
The distributions $\delta q^f$ are related to the 
transversity functions $h_1^f$:

\begin{equation}
h_1^f(x) = \int d^2 p_{\perp} \delta q^f (x, {\bf p}_{\perp}),
\label{rh1}
\end{equation}
a similar relation holding true for antiquarks.

Now we discuss a symmetry property of the spin density functions $\delta q^f$. 
Invariance of strong interactions under parity, time reversal and rotations (in 
particular rotations of $\pi$ around the proton momentum) imply

\begin{equation}
\delta q^f (x, {\bf p}_{\perp}) = \delta q^f (x, -{\bf p}_{\perp}).
\label{symde}
\end{equation}
However, if we take into account the initial state interactions and soft gluon 
exchange between the protons, we have to introduce also the {\it 
effective}, T-odd, density functions\cite{an1}, which do not fulfil eq. 
(\ref{symde}), since time reversal invariance does not apply trivially to them. 
This fact has no particular 
consequences on the integral (\ref{dqf}), which in general does not vanish for 
$|{\bf q_{\perp}}|$ of order 1 $GeV$. However, if we integrate the cross 
section over the transverse momentum of the muon pair, the the antisymmetric 
hadronic tensor derives its contribution (if any) from the sole 
{\it effective} density functions. 

Inserting eq. (\ref{ht2}) into  eq. (\ref{ht0}), we get 

\begin{equation}
H^{(0)}_{\mu\nu} = S^{(0)}_{\mu\nu} + i A^{(0)}_{\mu\nu},
\label{ht3}
\end{equation}
where

\begin{equation}
S^{(0)}_{\mu\nu} = \sum_{f=1}^3 e^2_f S^f_{\mu\nu},  \ ~~~~~~~~ \ ~~~~~~~~~
A^{(0)}_{\mu\nu} = \sum_{f=1}^3 e^2_f A^f_{\mu\nu}.
\label{ht33}
\end{equation} 
In order to find the asymmetries $A_1$ and $A_2$ (see eqs. (\ref{as1}) and 
(\ref{as2})), we combine the hadronic tensor 
(\ref{ht3}) with the leptonic tensor (\ref{lept1}), according to formula 
(\ref{dsg}) for the differential cross section. 
Therefore, in the zero order approximation
of the QCD  parton model, we find the following results:

$~i)$ If the helicity of the final negative muon is 
detected, the leptonic tensor has a nonvanishing antisymmetric part, 
which, combined with the antisymmetric part of the hadronic tensor (\ref{ht3}), 
gives a nonzero contribution to the muon helicity
asymmetry $A_1$ .

$ii)$ On the contrary, if no helicity is detected, only the symmetric part of 
the leptonic tensor survives and combination with the hadronic tensor drops out
$A^{(0)}_{\mu\nu}$. Since $S_{\mu\nu}^f$ (see eq. (\ref{symte})) is spin 
independent, the left-right asymmetry $A_2$ vanishes.
This result may be also deduced from parity and time reversal invariance.

Concerning the muon helicity asymmetry, we are faced with two important 
questions:

1) The asymmetry $A_1$ and the transversity function, $h_1^f$, depend on the 
distribution function $\delta q^f$ through integral relations, respectively  
eq. (\ref{dqf}) and eq. (\ref{rh1}). Therefore, we have to prove that the 
function we extract from eq. (\ref{dqf}) contributes to $h_1^f$.

2)  One may wonder where a does chiral-odd function like $h_1^f$ come from in 
the process considered. Such a distribution function arises, {\it e. g.,} in DY 
with two polarized proton beams, or from a mass term, which causes helicity 
flip.

To answer the first question, we write the relation

\begin{equation}
\delta q^f = cos \alpha ~ (\delta q^f_R - \delta q^f_L) + 
sin \alpha ~ \delta q^f_H,
\label{decp}
\end{equation}
where
\begin{eqnarray}
\delta q^f_{R(L)} &=& |<q_{R(L)}|P^{\uparrow}>|^2,
\\
\delta q^f_R &=& 2 Re [<P^{\uparrow}|q_R><q_L|P^{\uparrow}>],
\\
cos \alpha &=& \frac{{\bf p}_T\cdot{\bf S}}{xP}. \ ~~~~~~~ ~~~~~ ~~~~ \
\end{eqnarray}

Here $|P^{\uparrow}>$ and $|q_{R(L)}>$ are shorthand notations for a canonical 
state of the proton and for quark helicity states. Relation (\ref{decp}) 
follows from 
decomposing into helicity states a quark state with a given transverse momentum 
and spin parallel or antiparallel to the proton spin. The first term of that 
relation, which is twist 3 and chiral even, may be expressed as a 
linear combination of the usual helicity distribution functions. On the 
contrary, the second term is chiral odd and prevalently twist 2, since $sin 
\alpha$ $\sim$ 1. Thanks to arguments similar to those which have led to 
relation (\ref{symde}), we conclude that the first term of eq. (\ref{decp}) 
does not contribute to $h_1^f$, as can be seen from eq. (\ref{rh1}). The 
transversity function depends solely on the chiral-odd function $\delta q^f_H$, 
which contributes to the muon helicity asymmetry, as can be checked by 
substituting eq. (\ref{decp}) into eq. (\ref{dqf}).

As regards the origin of the chiral-odd function in our process, we show in 
appendix A that the antisymmetric tensor (\ref{asmte}) may be written as
\begin{equation}
A^f_{\mu\nu} = m_f \epsilon_{\alpha\mu\beta\nu} \int d^2 
p_{1\perp}\sum_{T_1,T_2}
\left[q^f_{T_1}(x_1,{\bf p}_{1\perp})\bar{q}^f_{T_2}(x_2,{\bf p}_{2\perp}^2) 
S_f^{T_1\alpha}p_1^{\beta}-(1\leftrightarrow 2)\right].
\label{mas}
\end{equation}
Here $m_f$ and $S_f^{T_1}$ are respectively the mass and the Pauli-Lubanski 
four-vector of the active quark. The quark mass causes helicity flip and 
therefore a contribution to the chiral-odd distribution function. The smallness 
of $m_f$ is compensated by $S_f^{T_1}$, which results to be (see Appendix A) 
\begin{equation}
S_f^{T_1} \sim 2T_1 {\bf p}_{1\perp}\cdot {\bf S} \frac{\sqrt 2 n_1}{m_fx_1}.
\end{equation}

The effect we have just illustrated is washed out in DIS, since we have to 
convolute the elementary quark-photon cross section over only one distribution 
function. In this case also the initial state interactions, and therefore the 
{\it effective} distribution functions, are suppressed.

\section{QCD first-order contributions to spin \\
asymmetries}

Now we consider the $q-{\bar q}$ annihilation amplitudes with one "real" gluon 
emitted (absorbed) by one of the colliding hadrons and absorbed (emitted) 
by the active parton of the other 
hadron. These interfere with the amplitudes just considered in the preceding 
section. The interference terms are twist-three 
and first order in $g$. We adopt the axial gauge $A^a\cdot n_2 = 0$, where 
$A^a$ is the gluonic field, $a = 1...8$ and $n_2$ is given by eq. 
(\ref{verso}). This gauge is not covariant, but 
allows to keep the parton model description; in particular, in our
approximation, this gauge 
avoids the complication of "link" operators\cite{bo3}, which we should introduce
in a generic gauge, in order to ensure gauge invariance of the nonperturbative 
("soft") functions involved in scattering. 

In this section we write the twist-three contributions according to the parton 
model\cite{efp}; secondly we impose gauge invariance and deduce symmetry 
properties of the "soft" functions that we are going to introduce, the so-called 
correlation functions; thirdly we discuss the behavior of the fermionic 
propagator, which appears in the "hard" coefficient; lastly we perform 
calculations.

\subsection{Parton Model Approach}

The twist-three contribution to the hadronic tensor reads, in tree 
approximation,

\begin{equation}
H'_{\mu\nu} = g \sum_{f=1}^3 e_f^2 (H_{\mu\nu}^{'f,a} - H_{\mu\nu}^{'f,b}),
\label{htt3}
\end{equation}
where
\begin{equation}
H_{\mu\nu}^{'f,a} = \sum_{l=1}^2H_{\mu\nu}^{'f,a,l}, \ ~~~~~~ \  ~~~~~~~ \ 
~~~~~~ \ H_{\mu\nu}^{'f,b} = \sum_{l=1}^2H_{\mu\nu}^{'f,b,l}. \label{htt4}
\end{equation}
Each term in eqs. (\ref{htt4}) corresponds to a different graph:

$~i)$ $H_{\mu\nu}^{'f,a,l}$ ($l=1,2$) refer to a gluon emitted (or absorbed) by 
proton 1 - which is polarized - and absorbed (or emitted) by the active 
antiquark ($l$ = 1) or quark ($l$ = 2) of proton 2.

$ii)$ $H_{\mu\nu}^{'f,b,l}$ refer instead to a gluon emitted (or 
absorbed) by proton 2 and absorbed (or emitted) by the active parton of proton 
1.
Analytically we have  
\begin{eqnarray}
H_{\mu\nu}^{'f,a,1} = \int d\Omega_1 \sum_{T_1,T'_1,T_2}
c^{f,1}_{T_1, T'_1} (x_1,{\bf p}_{1\perp}; x'_1, 
{\bf p'}_{1\perp}) {\overline q}^{f,2}_{T_2} (x_2, 
{\bf p}_{2\perp}) h^{'{\overline T}_{12}}_{\mu\nu} 
(x_1, x'_1, x_2), 
\label{hgt1} 
\\
H_{\mu\nu}^{'f,a,2} = \int d\Omega_1 \sum_{T_1,T'_1,T_2}
{\bar c}^{f,1}_{T_1, T'_1} (x_1, {\bf p}_{1\perp}; x'_1, 
\ {\bf p'}_{1\perp}) q^{f,2}_{T_2} (x_2, 
{\bf p}_{2\perp}) {\bar h'}^{{\overline T}_{12}}_{\mu\nu} 
(x_1, x'_1, x_2),
\label{hgt2} 
\end{eqnarray}
 while $H_{\mu\nu}^{'f,b,l}$ are obtained from (\ref{hgt1}) and (\ref{hgt2}) by 
substituting $(1\leftrightarrow 2)$ and $\mu\leftrightarrow\nu$.
$c^{f,l}_{T_l, T'_l}$ $({\overline c}^{f,l}_{T_l, T'_l})$ ($l$ = 1, 2) is the 
correlation function between a quark (antiquark) of four-momentum $p_l$ and 
spin $T_l$ and another of four-momentum $p'_l$ and spin $T'_l$.  
$d\Omega_l$ is the phase-space element, {\it i. e.},

\begin{eqnarray}
d\Omega_l &=& d\Gamma_q \frac{1}{(2\pi)^3} d^4 k_l \delta (k_l^2) d^4 p'_l 
\delta (p_l^{'2}) \delta^4 (p_l - p'_l - k_l) \theta(p'_{l0}) 
\nonumber
\\
&\to&  \frac{1}{2(2\pi)^5 Q^2} d^2p_{l\perp} d^2k_{l\perp} \frac{dz_l}{z_l},
\label{ps1}
\end{eqnarray}
where $d\Gamma_q$ is given by formula (\ref{qps}), $k_l$ is the four-momentum 
of the absorbed (or emitted) gluon and $z_l$ = $x_l'-x_l$. The arrow denotes 
integration over $p_{l0}$, $p_{l\parallel}$, $k_{l0}$, $k_{l\parallel}$. Lastly
in the axial gauge $A^a\cdot n_2 = 0$ we have

\begin{equation}
h^{'{\overline T_{12}}}_{\mu\nu} = \frac{1}{3} \frac{1}{{\bar 
p}_2^2+i\epsilon} Tr \left[\kappa^{T_1, T'_1} \gamma_{\mu} 
\left(\rlap/{\bar p}_2 \rlap/e {\bar \rho}^{T_2} + {\bar \rho}^{T_2} \rlap/e 
{\rlap/{\bar p}_2}\right)\gamma_{\nu}\right],
\label{hppp}
\end{equation}
where
\begin{equation}
 {\overline T_{12}} \equiv (T_1, T'_1,T_2), \ ~~~~ \ ~~ {\bar p}_2 = p_2-k_1,
 \label{pol} 
\end{equation} 
$e_{\mu}$ is the polarization four-vector of the gluon and $\kappa^{T, T'}$ 
the spin correlation matrix between a quark of spin $T$ and 
another with spin $T'$. ${\bar h}^{'{\overline T_{12}}}_{\mu\nu}$ is obtained 
from eq. (\ref{hppp}) by changing the quark matrices with the antiquark 
matrices and vice-versa, {\it i. e.}, $\kappa^{T_1, T'_1}$ $\to$ 
$\bar{\kappa}^{T_1, T'_1}$ and 
${\bar \rho}^{T_2}$ $\to$ $\rho^{T_2}$. As shown in Appendix B, 
\begin{equation}
\kappa^{T, T'} =\psi(t) \kappa_p^{T, T'}, \ ~~~~~ 
\bar{\kappa}^{T, T'} =\psi(t) \bar{\kappa}_p^{T, T'}, \label{corr03}
\end{equation} 
where
\begin{eqnarray}
t &=& \frac{x-x'}{x+x'},\ ~~~~ \ ~~~~ \ ~~~~ 
\ ~~~~ \ ~~~~ \ ~~~~ \psi(t) = \frac{1+t}{\sqrt{1-t^2}},
\label{tpsi}
\\
\kappa_p^{T, T} &=& {\bar \kappa}_p^{T, T} = 
\frac{1}{2}\rlap/p(1+2T\gamma_5 \rlap/S), ~~~~~~~~~
\kappa_p^{T,-T} ({\bar \kappa}_p^{T,-T}) = \frac{1}{2}\rlap/p \gamma_5 
(\rlap/{\bar S}\mp 2iT). \label{corr99}
\end{eqnarray}
Since we are considering the twist-three contribution, we neglect the 
transverse momentum in the numerators of
$h^{'{\overline T}}_{\mu\nu}$ and ${\bar h}^{'{\overline T}}_{\mu\nu}$. In 
this approximation the spin density matrix reads

\begin{equation}
\rho^T = {\overline \rho}^T = \frac{1}{2} \rlap/p(1 + 2T \gamma_5 \rlap/S).
\end{equation}

In the next subsection we shall show that the sum  
$H^{(0)}_{\mu\nu} + H'_{\mu\nu}$ 
- where $H^{(0)}_{\mu\nu}$ is given by eq. (\ref{ht0}) and $H'_{\mu\nu}$ by
eq. (\ref{htt3})) - is gauge invariant, although the single terms are not.

Since the proton 1 is transversely polarized, quark-gluon interactions
involve linearly polarized gluons\cite{la}, either in the 
direction of the spin of the proton, or in the direction orthogonal to the 
spin and to the momentum of the proton. In Appendix B we show that

\begin{eqnarray}
T'_1 = T_1: ~~ e = 2T_2{\bar S}; \ ~~~~~~~~ \ ~~~~~~~~ \ ~~~~~~~~ \ ~~~~~~~~  
T'_2 = T_2: ~~ e = 2T_1{\bar S}; \label{gl100}
\\
T'_l \neq T_l: ~~ e = -S\delta_{T_1,T_2}. \ ~~~~~~~~ \ ~~~~~~~~ \ ~~~~~~~~ \ ~~
\label{gl200}  
\end{eqnarray}

Substituting eqs. (\ref{corr03}) and (\ref{corr99}) to (\ref{gl200}) into eq. 
(\ref{hppp}), and developing calculations, we get
 
\begin{equation}
h^{'{\overline T_{12}}}_{\mu\nu} = 
h^{'\tilde{T}_{12}}_{N\mu\nu} \delta_{T'_1, T_1} 
+ h^{'\tilde{T}_{12}}_{F\mu\nu} \delta_{T'_1,-T_1},
\label{qt1}
\end{equation}
where $\tilde{T}_{12}$ $\equiv$ $(T_1, T_2)$ and
  
\begin{eqnarray}
h^{'\tilde{T}_{12}}_{N\mu\nu} &=&\frac{1}{3} \psi (t_1)
\frac{2p_2\cdot k_1}{{\bar p}^2_2+i\epsilon} 2(T_1+T_2)
(p_{1\mu}{\bar S}_{\nu}+p_{1\nu}{\bar S}_{\mu}),
\label{sus1}
\\
h^{'\tilde{T}_{12}}_{F\mu\nu} &=&\frac{1}{3} \psi (t_1)
\frac{2p_2\cdot k_1}{{\bar p}^2_2+i\epsilon} \left[ 2T_1 
\epsilon_{\alpha\mu\beta\nu} S^{\alpha}p_1^{\beta}+ 
2T_2(p_{1\mu}{\bar S}_{\nu}+p_{1\nu}{\bar S}_{\mu})\right]\delta_{T_1,T_2}.
\label{sus2}
\end{eqnarray}
Here the suffixes $N$ and $F$ denote repectively spin nonflip and flip of the 
active quark. 
The tensors ${\bar h'}^{\tilde{T}_{12}}_{N(F)\mu\nu}$ are equal to expressions
(\ref{sus1}) - (\ref{sus2}), except for a change of sign in front of the first 
term of eq. (\ref{sus2}). $h^{'\tilde{T}_{21}}_{N(F)\mu\nu}$ and 
${\bar h'}^{\tilde{T}_{21}}_{N(F)\mu\nu}$,  involved in the expressions of
$H_{\mu\nu}^{'f,b,l}$, are obtained from $h^{'\tilde{T}_{12}}_{N(F)\mu\nu}$ and 
${\bar h'}^{\tilde{T}_{12}}_{N(F)\mu\nu}$ by 
substituting $(1\leftrightarrow 2)$ and $\mu\leftrightarrow\nu$. 
Now we establish some important relations, which considerably simplify our 
calculations.

\subsection{Correlation Functions: Normalization and \\
Symmetry Properties}
\label{subsec:sym}

In Appendix C we show that gauge invariance implies the following 
normalization for the quark correlation functions:

\begin{eqnarray}
c^{f,l}_{T_l,T_l'} (x_l, {\bf p}_{l\perp}; x'_l, 
{\bf p}'_{l\perp}) \delta^4 (p_l - k_l - p'_l) = 8\pi^{3/2} (|{\bf p}_l| 
|{\bf k}_l| |{\bf p'}_l|)^{1/2}\nonumber\\
\times \sum_{i,j=1}^3 \sum_{c=1}^8<P_l|c^{f\dagger}_{T'_l,i} 
({\bf p'}_l) a^c_m({\bf k}_l) 
c^f_{T_l,j} ({\bf p}_l) |P_l>.
\label{cort}
\end{eqnarray}
Here $i$, $j$ and $c$ are the color indices of the quarks and of 
the gluon, combined in such a way that the operator product 
constitutes a color singlet. $c^f_{T,i}$ and $c^{f\dagger}_{T,i}$ are the 
destruction and creation operators for the quarks,
$a^c_m({\bf k}_l)$ $(m = 1, 2)$ are destruction operators for 
gluons whose polarization four-vector is $S$ ($m = 1$, 
corresponding to $T'_l \neq T_l$), or 
$\bar{S}$ ($m = 2$, corresponding to $T'_l = T_l$). $a^c_m({\bf k}_l)$    
has to be substituted by $a^{c\dagger}_m({\bf k}_l)$ for negative values of 
the gluon energy.

Invariance under parity inversion, time reversal, and rotation by $\pi$ around 
the proton momentum imply, through  eq. (\ref{cort}),

\begin{equation}
c^{f,l}_{T'_l, T_l} (x'_l, {\bf p}'_{l\perp}; x_l, {\bf p}_{l\perp}) = 
- c^{f,l}_{T_l, T'_l}  (x_l, -{\bf p}_{l\perp}; x'_l, -{\bf p}'_{l\perp}). 
\label{c1}
\end{equation}
To show that, we observe that the combined action of the three transformations 
leaves spin and longitudinal momenta unchanged, while inverting the arguments 
of the function and transverse momenta. Moreover time reversal produces the 
same change of phase ($n\pi$, with $n$ integer) as the rotation. Therefore the 
minus sign in front the r.h.s. of (\ref{c1}) is due to the product of the 
intrinsic parities of the two fermions and of the gluon.

Furthermore the hermiticity condition yields
\begin{equation}
c^{f,l}_{T'_l, T_l} (x'_l, {\bf p}'_{l\perp}; x_l, {\bf p}_{l\perp}) = 
c^{f,l*}_{T_l, T'_l} (x_l, {\bf p}_{l\perp}; x'_l, {\bf p}'_{l\perp}). 
\label{c2}
\end{equation}
Therefore, if we integrate over transverse momenta, the correlation functions
are imaginary and antisymmetric under 
the simultaneous exchange of the momenta and spin of the two quarks. 
Relations similar to eqs. (\ref{cort}) - (\ref{c2}) hold true for the antiquark 
correlation functions. Lastly, since the proton 2 is unpolarized, we have 
\begin{equation}
c^{f,2}_{T_2, T_2'} = c^{f,2}_{-T_2, -T_2'}, \ ~~~~~~~~ \ ~~~~~~~~ \ ~~~
{\bar c}^{f,2}_{T_2, T_2'} = {\bar c}^{f,2}_{-T_2, -T_2'}.
\label{c3}
\end{equation}
These equations, together with eqs. (\ref{rel2}), imply that the terms 
proportional to 
$T_2$ in eqs. (\ref{sus1}) and (\ref{sus2}) (or in those which are obtained by 
substituting $(1\leftrightarrow 2)$) are dropped after summing over the spin 
indices.

The symmetry property (\ref{c1}) does not hold if we take into account
the initial state interaction and soft gluon exchange between the protons. 
Indeed, analogously to the {\it effective} density functions, it makes sense to 
introduce {\it effective}, T-odd, correlation functions, for which time 
reversal invariance cannot be expressed in a trivial way\cite{an1}. Therefore 
such functions - whose importance will be shown in the following subsection - 
have in general a real part even after integration over 
transverse momentum.

\subsection{The fermionic propagator}
In twist-3 approximation the tensor (\ref{hppp}) - the "hard" coefficient of 
the hadronic tensor - is independent of the transverse momentum of the active 
parton. Therefore our preceding considerations imply that, aside from the
{\it effective} correlation functions, the left-right asymmetry of the cross 
section {\it integrated} over the transverse 
momentum of the muon pair receives a contribution from first order 
corrections, only if the tensor (\ref{hppp}) has an 
imaginary part. This is in agreement with the observation by Boer et 
al.\cite{bo1}, although we arrive at a 
different conclusion. We examine the question in detail.

The tensor (\ref{hppp}) includes the factor $(2k_1\cdot 
p_2)/(\bar{p}_2^2+i\epsilon)$.
For a massless quark $\bar{p}_2^2 = -2k_1\cdot 
p_2$, which annihilates the effects of the imaginary part of the tensor, unless 
the correlation function $c^{f,l}_{T_l,T_l'}$ has a simple pole - a {\it 
gluonic pole}\cite{bo1} - just at $\bar{p}_2^2$ = 0. If we would neglect, like
Boer et al.\cite{bo1}, the transverse momentum effects in the fraction above, 
$\bar{p}_2^2$ would vanish at $z_1 = 0$ and it would make sense to assume a 
{\it gluonic pole} just at zero momentum  
(a soft gluon contribution), so that the fermion propagator would 
yield an imaginary part. But, at least in the approximation of on-shell and 
massless quarks - adopted also by Boer et al.\cite{bo1} -, the transverse 
momentum has dramatic consequences on the pole, which results to be located at 
\begin{equation} 
z_1 = \frac{x_2P{\bf p}_{2 \perp}\cdot{\bf k}_{1 \perp}\pm\sqrt{\Delta}}
{P{\bf p}_{2 \perp}^2},
\ ~~~~~~ \ ~~~~~ \
\Delta = (x_2^2 P^2 + {\bf p}_{2\perp}^2) [({\bf p}_{2 \perp}\cdot{\bf k}_{1 
\perp})^2 - {\bf p}_{2 \perp}^2{\bf k}_{1 \perp}^2].
\label{eqq}
\end{equation} 
This value is generally complex, and by no means can be approximated by 
$z_1$ $\simeq$ 0. Consequently a soft {\it gluonic pole} would not imply an 
imaginary part for the tensor (\ref{hppp}). On the other hand it appears quite 
arbitrary to assume that $c_{T T'}^{f,2}$ has a pole located just in 
correspondence of the value given by the first eq. (\ref{eqq}). Analogous 
considerations can be done for the pole $(\bar{p}_1^2)^{-1}$, with $\bar{p}_1 = 
p_1 - k_2$. Therefore the "hard" coefficient of the hadronic tensor has not 
an imaginary part. We just set the 
fractions in front of the tensors (\ref{sus1}) and (\ref{sus2}) equal to -1. 
As a consequence the 
twist-3 contribution (if any) to the left-right asymmetry of the {\it 
integrated} cross section is entirely due to the 
{\it effective} correlation functions. Our considerations are somewhat 
analogous to 
those of Anselmino et al.\cite{an1}, concerning pion inclusive production.

\subsection{Calculation of the gluon correction}

Now we perform the calculation of the hadronic tensor (\ref{htt3}). 
Substituting eqs. (\ref{hgt1}) to (\ref{ps1}) into eqs. 
(\ref{htt4}), and taking into account the results (\ref{sus1}) and (\ref{sus2}) 
and the considerations above, we get

\begin{eqnarray}
H_{\mu\nu}^{'f,a} &=& \frac{\sqrt{2} x_1P}{6 (2\pi)^5Q^2} 
(s_{1\mu\nu} \tilde{\delta}_1 Q^f + i a_{1\mu\nu} \hat{\delta}_1 Q^f ),
\label{hha}
\\
H_{\mu\nu}^{'f,b} &=&  \frac{\sqrt{2} x_2P}{6 (2\pi)^5Q^2} 
(s_{2\mu\nu} \tilde{\delta}_2 Q^f + i a_{2\mu\nu} \hat{\delta}_1 Q^f). 
\label{hhb}
\end{eqnarray}

Here we have set
\begin{eqnarray}
\tilde{\delta}_1 Q^f &=& \int d^2 p_{1\perp} [\delta C^f_{S,1} (x_1, 
{\bf p}_{1\perp}) {\overline q}^f_2 (x_2, {\bf p}^2_{2\perp}) + \delta
{\overline C}^f_{S,1} (x_1, {\bf p}_{1\perp}) q^f_2 (x_2, {\bf p}^2_{2\perp})],
\label{lras}
\\
\tilde{\delta}_2 Q^f &=& \int d^2 p_{1\perp} [C^f_{S,2} (x_2, {\bf p}_{2\perp}) 
\delta{\overline q}^f_1 (x_1, {\bf p}_{1\perp}) + {\overline C}^f_{S,2} 
(x_2,{\bf p}_{2\perp}) \delta q^f_1 (x_1, {\bf p}_{1\perp})],
\label{lras2}
\\
\hat{\delta}_1 Q^f &=& \int d^2 p_{1\perp} [\delta C^f_{A,1} (x_1, 
{\bf p}_{1\perp}) {\overline q}^f_2 (x_2, {\bf p}^2_{2\perp}) - \delta
{\overline C}^f_{A,1} (x_1, {\bf p}_{1\perp}) q^f_2 (x_2, {\bf p}^2_{2\perp})],
\label{hat1}
\\
\hat{\delta}_2 Q^f &=& \int d^2 p_{1\perp} [C^f_{A,2} (x_2, {\bf p}_{2\perp}) 
\delta{\overline q}^f_1 (x_1, {\bf p}_{1\perp}) - {\overline C}^f_{A,2} 
(x_2,{\bf p}_{2\perp}) \delta q^f_1 (x_1, {\bf p}_{1\perp})]
\label{hat2}
\end{eqnarray}
and
\begin{equation}
s_{l\mu\nu} = {\overline S}_{\mu} n_{l \nu} + {\overline S}_{\nu}  n_{l \mu},
\ ~~~~~~ \ ~~~~~~ 
a_{l\mu\nu} = \epsilon_{\alpha\mu\beta\nu}S^{\alpha}n_l^{\beta}, \ ~~~~~~ \ ~~~~~~ 
(l=1,2).
\end{equation}
 
Moreover 
\begin{eqnarray}
\delta C^f_{S,1} (x_1, {\bf p}_{1 \perp}) &=& \sum_T 2T C^{f,1}_{S,T} (x_1, 
{\bf p}_{1 \perp}), \label{int11}
\\
C^f_{S,2} (x_2, {\bf p}_{2 \perp}) &=& \sum_T  C^{f,2}_{S,T} (x_2, 
{\bf p}_{2 \perp}),
\label{int12}
\\
\delta C^f_{A,1} (x_1, {\bf p}_{1 \perp}) &=& \sum_T 2T C^{f,1}_{A,T} (x_1, 
{\bf p}_{1 \perp}), \label{int21}
\\
C^f_{A,2} (x_2, {\bf p}_{2 \perp}) &=& \sum_T  C^{f,2}_{A,T} (x_2, 
{\bf p}_{2 \perp}),
\label{int22}
\end{eqnarray}

where

\begin{eqnarray}
C^{f,l}_{S,T} ((x_l, {\bf p}_{l \perp})  &=&
Re \left\{\int d^2 k_{l\perp} \int 
\frac{dz_l}{z_l} \left[ \sum_{T'} c^{f,l}_{T, T'} 
(x_l, {\bf p}_{l \perp}; x'_l, {\bf p}'_{l\perp}) \right] \psi(t_l) \right\},
\label{int33}
\\
C^{f,l}_{A,T} ((x_l, {\bf p}_{l \perp})  &=&
Im \left\{\int d^2 k_{l\perp} \int 
\frac{dz_l}{z_l} \left[ c^{f,l}_{T, -T} 
(x_l, {\bf p}_{l \perp}; x'_l, {\bf p}'_{l\perp}) \right] \psi(t_l) \right\}
\label{int44}
\end{eqnarray}

and $l = 1,2$. Analogous expressions hold for the barred quantities. 
Here $z$ and 
${\bf k}_{\perp}$ are defined in the expression (\ref{ps1}) of the phase-space 
element, while $t$ is given by the first eq. (\ref{tpsi}). The QCD 
first order correction $H'_{\mu\nu}$ is obtained substituting eqs. (\ref{hha}) 
and (\ref{hhb}) into eq. (\ref{htt3})). Three remarks are in order.
\begin{itemize}
\item The functions $C^f_S$, $\delta C^f_S$, $C^f_A$ and $\delta C^f_A$ have 
the dimensions of a mass, as can be checked from eq. (\ref{cort}). 

\item Integrating  the cross section over the transverse momentum of the muon 
pair amounts to independently integrating over the quark transverse momentum 
the "soft" functions involved in the hadronic tensor. In particular, as stressed
in the previous subsection, the integrals of the functions $\tilde{\delta}_l 
Q^f$ ($l = 1,2$, see eqs. (\ref{lras}) and (\ref{lras2})) derive their 
contributions from the sole {\it effective} correlation functions.

\item Taking into account eqs. (\ref{tpsi}), the integrals (\ref{int33}) and
(\ref{int44}) can be split into two terms, corresponding to
\begin{equation}
\frac{1}{2} [{\tilde c}^f (x, x+z) +{\tilde c}^f (x, x-z)] \ ~~~~~ \ {\mathrm 
and} \ ~~~~~ \ \frac{1}{2z} [c^f (x, x+z) - c^f (x, x-z)],
\label{can}
\end{equation}
where $c^f = c^{f,l}_{T,T'} (1-t^2)^{-1/2}$, ${\tilde c}^f$ = $t/z 
c^f$ and $t$ is given by the first eq. (\ref{tpsi}). If we assume $c^f (x,x')$ 
= $c^f_0 (x,x')\delta(x-x')$, where $c_0^f$ is a smooth function of its 
arguments, the second term (\ref{can}) corresponds to the 
{\it derivative} term by Qiu and Sterman\cite{qi1,qi2}. (See ref.\cite{ha} for 
details). 
\end{itemize}

\section{Asymmetries}

In QCD first order approximation the hadronic tensor $H_{\mu\nu}$ consists,
according to eq. (\ref{hhtt}), of the sum of two terms, which have been
calculated in the two previous sections (formulae (\ref{ht3}) and  
(\ref{htt3})). 
Substituting the expressions of $H_{\mu\nu}$ and of the leptonic tensor
(\ref{lept1}) into the differential  cross section (\ref{dsg}), and taking into
account the formulae (\ref{as1}) and  (\ref{as2}) of the two asymmetries, we get

\begin{equation}
A_1 = A_1^{(0)} + A_1^{(1)},
\end{equation}
where
\begin{equation}
A_1^{(0)} =
\frac{8\sqrt{x_1x_2}}{Q} {{\sum_f e^2_f \delta Q^f} \over {\sum_f e^2_f 
\ Q^f}} {{cos \theta} \over {1+cos^2 \theta}} 
\label{helasy}
\end{equation}

and

\begin{equation}
A^{(1)}_1 =  {{-g}\over{(2\pi)^3\sqrt{x_1x_2} Q}}
{{\sum_f e^2_f (x_1\hat{\delta}_1 Q^f + x_2\hat{\delta}_2 Q^f)} \over 
{\sum_f e^2_f Q^f}} {{2sin \theta sin \phi} \over {1+cos^2 \theta}}.
\label{corrasy}
\end{equation}

Moreover

\begin{equation}
A_2 =  {{g}\over{(2\pi)^3\sqrt{x_1x_2} Q}}
{{\sum_f e^2_f (x_1\tilde{\delta}_1 Q^f + x_2\tilde{\delta}_2 Q^f)} \over 
{\sum_f e^2_f Q^f}} {{sin 2\theta cos \phi} \over {1+cos^2 \theta}}.
\label{lrasym}
\end{equation}

Here $\theta$ is the polar angle and $\phi$ the azimuthal angle of the negative 
muon: we have assumed a reference frame in the center-of-mass system of the 
muon pair, whose $z$-axis is taken along the momentum of the polarized proton,
while the $x$-axis is along the space component of $\bar{S}$. Furthermore
$Q^f, \delta Q^f, \tilde{\delta}_l Q^f, \hat{\delta}_l Q^f$ ($l = 1, 2$) are 
given, respectively, by eqs. (\ref{qf}), (\ref{dqf}) and (\ref{lras}) to 
(\ref{hat2}).

The expression of $A_2$ turns out to coincide with the one by Boer et 
al.\cite{bo1,bo3}, provided 

- we integrate the cross section over the transverse momentum of the virtual 
photon;

- we take into account the following notation differences: $q(x) \to 
f_1(x)$, ~~ $\phi \to \Phi_{S_1}-\frac{\pi}{2}$,

- we identify 

\begin{eqnarray}
M_P \tilde{h}^f (x) = -g\frac{P}{(2\pi)^3 Q} 
\int d^2 p_{\perp} C_S^{f} (x, {\bf p}_{\perp}),
\\
M_P \tilde{f}^f_T (x) = -g\frac{P}{(2\pi)^3 Q} 
\int d^2 p_{\perp} \delta C_S^{f} (x, {\bf p}_{\perp}), 
\label{boero}
\end{eqnarray}
where $M_P$ is the proton rest mass. 

\subsection{Zero order approximation} 
As we have seen, the left-right asymmetry at zero order vanishes. In the same 
approximation the muon helicity asymmetry is sensitive to the 
${\bf p}_{\perp}$-dependent 
transversity distributions $\delta q^f(x_1, {\bf p}_{1\perp})$ and $\delta 
{\overline q}^f(x_1, {\bf p}_{1\perp})$, which can be used for calculating the 
transversity distributions $h_1^f$ and ${\overline h}_1^f$ (see eq. 
(\ref{rh1})), with $f$ = 1,2,3. Eq. (\ref{helasy}) is a linear 
combination of the six unknown functions. In order to extract them, we need 
other, independent combinations.
For example, considering Drell-Yan events from collisions between a polarized 
proton beam and, say, a pion, we get an expression of the asymmetry analogous 
to $A_1$ (eq. (\ref{helasy})), where $q^f$ and ${\overline q}^f$ are replaced 
respectively by the quark and antiquark density functions inside the pion.
It is worth observing that Drell-Yan offers, at least in principle, a variety 
of independent combinations. For example, we could consider collisions of 
transversely polarized protons on unpolarized protons, antiprotons, positive 
and negative pions and kaons. In this way we could obtain six independent 
asymmetries similar to formula (\ref{helasy}), from which (or from part of
them) we could extract the unknown distribution functions by a fit,
taking into account symmetry  properties like isospin invariance, or general 
constraints like Soffer's  inequality\cite{so}.

In this connection it  is worth recalling that in ref.\cite{st} it was
proposed, quite analogously to us, to extract the quark helicity distributions
from Drell-Yan produced in scattering on longitudinally polarized proton target
of beams of pions and unpolarized  protons. They obtain an asymmetry
formula similar to $A_1^{(0)}$ (eq. (\ref{helasy})), although this is a 
twist-two and not a twist-three effect. 

It is known that our procedure of convoluting the elementary cross section over 
the transverse momentum of the active quarks is a good approximation only
for sufficiently large 
transverse momenta\cite{mos}. But this is not a severe constraint, 
since, as we have seen, transverse momentum is essential for exhibiting 
helicity asymmetry. Eq. (\ref{helasy}) suggests that directions not too 
far from the forward and backward one are the most proper. 

In order to estimate the order of magnitude of the asymmetry $A_1$, we take
into account the HERMES results\cite{ef},  {\it i. e.},
$|\delta q^f|/q^f$ $\sim$ $|\delta {\overline q}^f|/{\overline q}^f$ $\sim$ 
$(50\pm 30)\%$. Moreover eq. (\ref{helasy}) implies $A_1$ to vanish for $x_1$ =
$x_2$. Therefore, in order to maximize the asymmetry, we should take $x_1$ as 
different as possible from $x_2$, without making $\delta Q^f$ and $Q^f$
too small. For $Q^2$ of order 10 $GeV^2$ and $s$ $\sim$ 
100 $GeV^2$, a good choice, consistent with eq. (\ref{krel}), is $x_1\sim 0.5$, 
$x_2\sim 0.2$, or, viceversa, $x_2\sim 0.5$, $x_1\sim 0.2$. The evaluation of 
the asymmetry is particularly complicated, because the integrand at the r.h.s.
of eq. (\ref{dqf}) is partly positive and partly negative.
We just give an upper limit: since $\bar{|{\bf p}_{\perp}|}$ $\sim$ 0.5 $GeV$, 
under the conditions illustrated above we estimate $|A_1|$ $\leq$ 
$(10 \pm 6)\%$. 

\subsection{First order corrections}

In order to find the order of magnitude of the first order correction to the 
two asymmetries, we generalize the  Qiu-Sterman\cite{qi1} guess:
\begin{equation}
C^f_{S,T} (x,{\bf p}_{\perp}) = K_S q^f_T(x, {\bf p}_{\perp}), \ ~~~~~~ \ ~~~~~~ 
\ ~~~~~~ C^f_{A,T} (x,{\bf p}_{\perp}) = K_A q^f_T(x, {\bf p}_{\perp}), 
\label{coe1}
\end{equation}
where $C^f_{S,T}$ and $C^f_{A,T}$ are given by eqs. (\ref{int33}) and 
(\ref{int44}) and $K_{S}$ and $K_A$ are constants and independent of $T$.
Then eqs. (\ref{int11}) to (\ref{int22}) imply
\begin{equation}
C_S^f (x,{\bf p}_{\perp}) = K_S q^f(x, {\bf p}^2_{\perp}), 
~~~~~~~ \ ~~~~~ \ \delta C^f_S (x,{\bf p}_{\perp}) = 
K_S \delta q^f(x, {\bf p}_{\perp}), \label{paramds}
\end{equation}
analogous relations holding true for $C_A^f$ and $\delta C_A^f$. Two remarks 
are in order.
\begin{itemize}
\item The generalization (\ref{coe1}) of the Qiu-Sterman model is quite natural 
and immediate in our approach, whereas it would be rather complicated in the 
formalism of quantum field theory. 

\item To be precise, we have generalized one of the two guesses proposed by
Qiu and Sterman, the other one being obtained by multiplying the 
r. h. s. of eqs. (\ref{paramds}) by $x$. The two models could be distinguished 
by examining data at very low $x$. 
\end{itemize}
According to refs.\cite{qi1,ha,gr}, we may set $g |K_S|$ = $0.08 
(2\pi)^3\sqrt{2}M_P$. On the other hand, since 
the sums (\ref{int11}) and (\ref{int12}) consist of four terms, whereas the 
sums (\ref{int21}) and (\ref{int22}) consist of two terms, in order to evaluate
the order of magnitude,
it seems not completely unreasonable to guess that $|K_A| \simeq 1/2 |K_S|$.
 
We fix some parameters as before: $Q^2$ $\sim$ 10 $GeV^2$, $s$ $\sim$ 100 
$GeV^2$, $x_1 = 0.5$ and $x_2=0.2$ (or viceversa). Moreover we take into 
account once more the above mentioned results by HERMES. Lastly, as regards the 
angular dependence, eqs. (\ref{lrasym}) and (\ref{corrasy}) suggest that the 
most favourable conditions for detecting the left-right asymmetry and the first
order correction to the muon helicity asymmetry are, respectively, $\theta$ 
$\sim$ $\phi$ $\sim$ $\pi/2$ and $\theta$ $\sim$ $\pi/4$, $3\pi/4$, $\phi$ = 
0, $\pi$. 
We find $|A_2|$ $\sim$ $(2\pm 1)\%$ and $|A_1^{(1)}|$ $\sim$ 
$(1\pm 0.6)\%$. The uncertainty on these two asymmetries could be even 
larger, if we 
take into account that the coupling constant $g$ is poorly known\cite{ha}.
It is worth noticing that the first order correction to the muon helicity 
asymmetry is much smaller than $A_1^{(0)}$, moreover it can be disentangled, 
since it exhibits a completely different angular dependence.

\vskip 0.30in

\section{Conclusions}

We have considered the muon helicity asymmetry ($A_1$) and the left-right 
asymmetry ($A_2$) in the Drell-Yan process $p p^{\uparrow} \to \mu^+ \mu^- X$.
The two quantities are measurable, although for $A_1$ some special care is 
required\cite{bol}. Let us recall the main results.

1) We have calculated the two asymmetries in tree approximation by means of 
the QCD improved parton model\cite{efp,qi2}, which amounts to replacing 
transverse momentum by a combination of momentum and gluon field, as dictated 
by the covariant derivative $D_{\alpha}$ (see formula (\ref{sof2})). Indeed 
both asymmetries vanish at the leading twist and turn out to be generated, at 
twist-three, by the transverse components of the covariant derivative, 
{\it i. e.},

\begin{displaymath}
D_{\alpha}S^{\alpha} \to A_1, \ ~~~~ \ ~~~~ \ 
\ D_{\alpha}{\overline S}^{\alpha} \to A_2. 
\end{displaymath}

In particular $A_{1}$ receives its main contribution from the first term of the 
covariant derivative, that is, from parton transverse momentum, while $A_{2}$ 
gets a contribution from the sole second term, corresponding to one 
gluon corrections. 

2) We have performed the calculation in the framework of parton model, assuming 
an axial gauge and introducing the correlation functions between two partons of 
different momenta in the hadron. We have matched this simple and intuitive 
formalism to the quantum field theory approach, obtaining, as a result, 
condition (\ref{cort}), which uniquely fixes the normalization of the 
correlation functions and simultaneously guarantees gauge invariance. This 
procedure is of quite general validity for twist-three terms, owing to the 
local character of the interaction.

3) If we integrate the cross section over the transverse momentum of the muon 
pair, only the {\it effective} density and correlation functions contribute to 
$A_1$ and $A_2$. Therefore it appears useful to consider these kinds of 
asymmetries as well as those from the differential cross section.

4) Our approach leads to a simple physical interpretation of the 
"soft" functions $C^f$ and $\delta C^f$ involved in the 
left-right asymmetry, and to a natural generalization of the 
Qiu-Sterman\cite{qi1} assumption, allowing for a quantitative evaluation of 
$A_2$. As a result, for standard values of $Q^2$ 
and $s$, this asymmetry is estimated to be a few percent. 

5) Under the same conditions the muon helicity asymmetry is probably a bit 
larger. This asymmetry is sensitive to the transverse momentum
dependent transversity distributions, $\delta q^f$ and $\delta {\bar q}^f$,
which are related to the usual transversity distributions through eq. 
(\ref{rh1}). 
One possible advantage of our method is that one can obtain independent combinations of 
such functions by performing various scattering experiments with beams of 
unpolarized protons, pions and kaons on a transversely polarized proton target.

{\bf Acknowledgments}:
{\it The author is deeply indebted to his friend prof. J. Soffer for useful and 
stimulating discussions on this work. Furthermore he is grateful to his friend 
prof. C. M. Becchi for important suggestions.}

\vskip 1.0in

\setcounter{equation}{0}
\renewcommand\theequation{A.\arabic{equation}}

\appendix{\large \bf Appendix A}

Here we calculate the antisymmetric part of the hadronic tensor, by inserting 
the mass of the quark in the density matrix. We show that it reduces to the 
expression (\ref{asmte}) in the limit of zero mass, the smallness of the mass 
being compensated by the "good" component of the Pauli-Lubanski (PL) 
four-vector of the active quark. The hadronic tensor reads
\begin{equation}
H_{\mu\nu}^{f} = \frac{1}{4\pi^2 Q^2}\int d^2 p_{1\perp} \sum_{T_1, T_2} 
[ q^{f}_{T_1} (x_1, {\bf p}_{1\perp}) 
{\overline q}^{f}_{T_2} 
(x_2, {\bf p}_{2\perp}) h^{\tilde{T}_{12}}_{\mu\nu} 
(x_1, x_2; S) + (1 \leftrightarrow 2)],
\label{hadt111}
\end{equation}
where
\begin{equation}
h^{\tilde{T}_{12}}_{\mu\nu} (x_1, x_2; S) = \frac{1}{3} Tr (\rho^{T_1} 
\gamma_{\mu} \bar{\rho}^{T_2}\gamma_{\nu}) 
\label{qut210}
\end{equation}
and
\begin{equation}
\rho^T (\bar{\rho}^T) = \frac{1}{2}(\rlap/p\pm m_f) (1+\gamma_5 \rlap/S^T_f).
\label{matd222}
\end{equation}
Here $p$ = $p_1$ or $p_2$ and the + and - sign refer, respectively, to the 
quark and to the antiquark; moreover $S^T_f$ is the PL four-vector 
of the quark. We have 
\begin{equation}
S^T_{f\alpha} = 2T L_{\alpha}^{\beta}\xi_{\beta}.
\label{pl322}
\end{equation}
Here $\xi \equiv (0,{\bf S})$ is the PL four-vector of the proton, in the frame 
where it is at rest. $L$ is the matrix of the boost which takes the quark from 
rest to a four-momentum $p \equiv (\sqrt{m_q^2+{\bf p}^2}, {\bf p})$, where 
${\bf p} \equiv ({\bf p}_{\perp}, xP)$ and P is the proton momentum. A standard 
calculation shows that 
\begin{equation}
S^T_f \simeq \frac{2T}{m_fx} {\bf p_{\perp}}\cdot {\bf S} (1,0,0,1).
\label{pl33}
\end{equation}
On the other hand the antisymmetric part of the tensor (\ref{hadt111}) turns 
out to be
\begin{equation}
A_{\mu\nu}^{f} = \frac{i}{4\pi^2 Q^2}\int d^2 p_{1\perp} \sum_{T_1, T_2} 
a_{\mu\nu}^f [ q^{f}_{T_1} (x_1, {\bf p}_{1\perp}) 
{\overline q}^{f}_{T_2} (x_2, {\bf p}_{2\perp}) - (1 \leftrightarrow 2)],
\label{asht99}
\end{equation}
where
\begin{equation}
a_{\mu\nu}^f = \frac{1}{3}m_f \epsilon_{\alpha\mu\beta\nu} S^{T_1\alpha}_f 
q^{\beta}
\end{equation}
and $q$ = $p_1 + p_2$ is the four-momentum of the virtual photon.
Taking into account formula (\ref{pl33}), we get
\begin{equation}
a_{\mu\nu}^f \simeq \frac{2T_1}{3} \frac{{\bf p}_{1\perp}\cdot {\bf S}}{x_1} 
\epsilon_{\alpha\mu\beta\nu} \sqrt{2} n_1^{\alpha} q^{\beta}.
\label{aqt67}
\end{equation}
Here $p_2 \simeq x_2 P \sqrt{2} n_2$ and $n_{1(2)} \equiv 
\frac{1}{\sqrt{2}}(1,0,0,\pm 1)$. Substituting the expression (\ref{aqt67}) 
into the antisymmetric tensor (\ref{asht99}), 
and recalling the kinematic relations deduced in sect. 3, $A_{\mu\nu}^{f}$ 
turns out to coincide with formula (\ref{asmte}).

\vskip 0.6in
\setcounter{equation}{0}
\renewcommand\theequation{B.\arabic{equation}}
\appendix{\large \bf Appendix B}

Here we derive the expression of the spin correlation matrix of two quarks, 
moreover we study in detail the  ~~~ coupling between a ~~~~ transversely 
polarized quark and a linearly polarized gluon.
\vskip 0.3in
\centerline{\bf B.1 -  Spin correlation matrix}
The spin correlation matrix is defined as

\begin{equation}
\kappa^{T, T'} (p, p') = u_T (p) {\overline u}_{T'} (p'),
\label{ron}
\end{equation}

where $u_T (p)$ is the Dirac spinor of a quark with four-momentum 
$p$ and third spin component $T$.
We consider a nonvanishing quark mass. First of all, we take both quarks at 
rest, secondly we make a Lorentz boost, thirdly we change the momentum of one 
of the two fermions; lastly we treat the limiting case of a negligible mass.

{\bf B.1.1} {\it - Both quarks at rest}  

Choosing, as usual, the $z$-axis as the quantization axis, we have

\begin{equation}
\kappa^{T, T'}_0 = \sum_{i,j = 1}^2 \chi^T_i \chi^{T'\dagger}_j a_{ij} =
a_{11} (1 + \sigma_3) + a_{12} (\sigma_1 + i \sigma_2) + a_{21} (\sigma_1 -
i \sigma_2) + a_{22} (1 - \sigma_3), 
\end{equation}
where $\chi^T$ is the usual spinor.
For pure spin states, such as those we are interseted in, some coefficients are 
nonvanishing: for $T$ = $T'$ $a_{12}$ and $a_{21}$ vanish, while for $T$ 
$\neq$ $T'$ $a_{11}$ = $a_{22}$ = 0. In particular we have
\begin{equation}
\kappa^{T, T}_0 = m (1+2 T \sigma_3), \ 
\ ~~~~~~~~ \ ~~~~~~~ \ 
\ \kappa^{T, -T}_0 = m (\sigma_1 + 2iT \sigma_2), 
\end{equation}
having choosen for the Dirac spinor the normalization ${\overline u}_{T} 
(p) u_T (p)$ = $2m$, where $m$ is the rest mass of the quark. The previous 
expressions can be generalized by introducing three mutually orthogonal unit 
vectors, ${\bf i}$, ${\bf j}$ and ${\bf k}$. Choosing the quantization axis 
along ${\bf k}$, we obtain

\begin{equation}
 \sigma_3 \rightarrow  \sigma_i  k^i, \ ~~~~~~ \ 
~~~~~ \sigma_1 +2 i T \sigma_2 \rightarrow  \sigma_i\ b_T^i, 
\ ~~~~~~ \ ~~~~~ \ {\bf b}_T = {\bf i} + 2iT {\bf j}.
\end{equation}
In terms of the Dirac matrices  the spin correlation matrix at rest can be 
written as

\begin{equation}
\kappa^{T, T}_0 = \frac{m}{2} (1\pm\gamma_0) (1-2 T \gamma_5 \gamma_i k^i), \ 
\ ~~~~~~~~  
\ \kappa^{T, -T}_0 = -\frac{m}{2} (1\pm\gamma_0) \gamma_5 
\ \gamma_i b_T^i, \nonumber
\end{equation}
where $m$ is the fermion rest ~~ mass and the ~~ $\pm$ sign ~~ in front 
of $\gamma_0$ ~~ refers ~~ to the (anti-)quark. 

{\bf B.1.2} {\it - Quarks with equal momenta}  

The previous formulae can be easily extended to the case of two quarks with
equal momenta. We get, similarly to the spin density matrix,

\begin{equation}
\kappa^{T, T}_p = \frac{1}{2} (\rlap/p \pm m) (1 + 2T \gamma_5 \rlap/k), 
\ ~~~~~~~ \ 
\kappa^{T, -T}_p = \frac{1}{2} (\rlap/p \pm m) \gamma_5 \rlap/b_T, 
\label{gem1}
\end{equation}
where, in the rest frame of the fermions, $k \equiv (0, {\bf k})$ and $b_T 
\equiv (0, {\bf b}_T)$.

It is worth noticing that, if the two quarks have different third spin 
components and momenta not parallel to the quantization axis, the final result 
(second formula (\ref{gem1})) depends on the choice of the $x$ and $y$-axis. We 
shall solve this ambiguity below, showing that the $y-z$ plane must be fixed in 
such a way to contain the quark momentum.

{\bf B.1.3} {\it - Quarks with different collinear momenta}  

Consider two quarks with different momenta, in a frame where they are collinear.
In this case the correlation matrix reads

\begin{equation}
\kappa^{T, T'} (p, p') = U(p, p') \kappa^{T, T'}_{p'}. \label{diff}
\end{equation}
Here $p$ $\equiv$ $(E, {\bf p})$ and $p'$ $\equiv$ $(E', {\bf p'})$ are the 
quark four-momenta in the frame considered; moreover $U(p, p')$ is the 
transformation matrix corresponding to the boost which changes $p'$ to $p$, 
{\it i. e.},
\begin{equation}
U(p, p') = \frac{1}{\sqrt{1-t^2}}\left(1+ \frac{p'_i}{|{\bf p}'|} 
\gamma_0\gamma_i t\right),
\end{equation}
where 
\begin{equation}
t = \frac{|{\bf p}'|-|{\bf p}|}{E'+E}.
\end{equation}
By using the Dirac equation, eq. (\ref{diff}) yields
\begin{equation}
\kappa^{T, T'} (p, p') = \frac{1}{\sqrt{1-t^2}}\left(1+ t
\frac{E'\mp\gamma_0m}{|{\bf p}'|}\right)\kappa^{T, T'}_{p'}. 
\label{coo6}
\end{equation}

{\bf B.1.4}{\it - Limit of zero quark mass}  

This limit is trivial in the case of $T'$ = $T$:
\begin{equation}
\kappa^{T, T} (p, p') = \frac{1+t}{2\sqrt{1-t^2}} 
\rlap/p'  (1 + 2T \gamma_5 \rlap/k). \label{ug}
\end{equation}
On the contrary, as we have observed above, if $T'$ $\neq$ $T$, 
and if the quark momenta ${\bf p}$ and
${\bf p}'$ are not parallel to the quantization axis, {\it i. e.} to ${\bf k}$,
we are faced with an ambiguity. In particular, let us consider the case when
${\bf p}$ is orthogonal to ${\bf k}$, which is of interest to us: the 
correlation matrix is different according as to whether we take ${\bf i}$ or 
${\bf j}$ along ${\bf p}$. We show that the latter choice is correct.

Take ${\bf j}$ along ${\bf p}$. Formula (\ref{coo6}) yields
\begin{equation}
\kappa^{T, -T} (p, p') = \frac{-1}{2\sqrt{1-t^2}}\left(1+ t
\frac{E'\mp\gamma_0m}{|{\bf p}'|}\right) (\rlap/p' \pm m) \gamma_5 
(2iT\rlap/p'_cm^{-1} + \rlap/l),  
\end{equation}
where $p'_c \equiv (|{\bf p}'|, 0, 0, E')$ and $l \equiv (0, {\bf i})$. In the 
limit of $m$ $\to$ 0, taking into account that $\rlap/p' \rlap/p'_c$ = 
$O(m^2/{\bf p}^{'2})$, we get
\begin{equation}
\kappa^{T, -T} (p, p') = \frac{1+t}{2\sqrt{1-t^2}} 
\rlap/p' \gamma_5 (\mp2iT + \rlap/l). \label{kap}
\end{equation}
In the specific case that we consider in sect. 4, we identify the four-vector
$l$ with $\bar{S}$, which is given by eq. (\ref{sbar}). In order to check our 
choice, consider two massless quarks with equal momenta. Taking the $y$-axis 
along the quark momentum, we have
\begin{equation}
u^{1/2} = \frac{1}{\sqrt{2}} (u_R + u_L), \ ~~~~~ \ ~~~~~ \ ~~~~~  
u^{-1/2} = \frac{i}{\sqrt{2}} (u_R - u_L),\label{kit}
\end{equation}
where $u_R$ and $u_L$ are respectively the spinors of the right-handed and of 
the left-handed quark. Substituting formulae (\ref{kit}) into eq. (\ref{ron}), 
we get
\begin{equation}
\kappa^{T,-T}_{p} = -\frac{i}{2} [2T(u_R{\bar u}_R - u_L{\bar u}_L) + 
u_L{\bar u}_R - u_R{\bar u}_L].\label{fre}
\end{equation}
The usual formulae for the density matrix\cite{la} yield $u_R{\bar u}_R - 
u_L{\bar u}_L$ = $\pm \rlap/p \gamma_5$. On the other hand, applying the 
second eq. (\ref{gem1}), we get $u_L{\bar u}_R - u_R{\bar u}_L$ = $i \rlap/p 
\gamma_5 \rlap/l$. Therefore eq. (\ref{fre}) turns out to coincide with eq.
(\ref{kap}) for $t$ = 0. Now return to eq. (\ref{coo6}) and take ${\bf i}$ 
(instead of ${\bf j}$) along ${\bf p}$. This choice yields a result which 
differs from eq. (\ref{kap}) by a factor $2iT$, and it does not  satisfy the 
self-consistency test that we have just illustrated.

Notice that in the massless case $t$ = $(x'-x)/(x'+x)$, where $x$ and $x'$ are 
the light cone fractional momenta of the quarks; therefore the expression 
(\ref{kap}) is covariant. We do not take into account the quark transverse 
momentum - which involves a Melosh-Wigner rotation -, however this is not 
requested in the twist-three approximation, which we are considering.

Lastly it is straightforward to see that in eqs. (\ref{ug}) and (\ref{kap}) we 
can interchange $p \leftrightarrow p'$. This property we exploit in subsect. 
4.1.

\vskip 0.30in
\centerline{\bf B.2 - Linearly polarized gluons}

Now we consider a gluon emitted by one of the two protons (say proton 2) and 
interacting perturbatively with the active quark of the other proton. In this 
connection we observe that linearly polarized gluons are naturally coupled to 
transversely polarized quarks. In the reference frame defined above, for a 
gluon travelling along the $y$-axis, we have\cite{la,ar2} 
\begin{equation}
G_x^{\mu} = \frac{1}{\sqrt{2}}(G_R^{\mu}+G_L^{\mu}), \ ~~~~~~ \ ~~~~~~~ \ 
~~~~~~~ G_z^{\mu} = \frac{i}{\sqrt{2}}(G_R^{\mu}-G_L^{\mu}), 
\label{gluu}
\end{equation}
where $G_x^{\mu}$ and $G_z^{\mu}$ denote gluon states polarized respectively 
along the $x$- and along the $z$-axis and $G_R$ and $G_L$ are right-handed and 
left-handed gluons. Colour indices have been omitted for the 
sake of simplicity. The perturbative coupling of a massless quark with a gluon 
is of the type
$G_R^{\mu}\bar{u}_R\gamma_{\mu}u_R$ or $G_L^{\mu}\bar{u}_L\gamma_{\mu}u_L$. On 
the other hand we have to do with quark states like (\ref{kit}), which couple 
with gluon states of the type (\ref{gluu}). Such couplings read

\begin{eqnarray}
({\bar u}^{T_1}\gamma_{\mu}u^{T_1})G_x^{\mu} &=& \frac{1}{2\sqrt{2}} 
(G_R^{\mu}\bar{u}_R\gamma_{\mu}u_R+G_L^{\mu}\bar{u}_L\gamma_{\mu}u_L),
\\
({\bar u}^{T_1}\gamma_{\mu}u^{-T_1})G_z^{\mu} &=& -\frac{2T_1}{2\sqrt{2}} 
(G_R^{\mu}\bar{u}_R\gamma_{\mu}u_R+G_L^{\mu}\bar{u}_L\gamma_{\mu}u_L).
\end{eqnarray}

Furthermore the gluon is related to the active quarks of protons 1 and 2 by 
angular momentum conservation, which implies that we have to take into account 
the factor $F$ = $sgn <1/2,T_1; 1,T_g |1/2,T'_1>$ = $2T_1$, where
$T_g = T'_2-T_2$ and $T'_1$ = $T_g + T_1$. Angular momentum conservation 
implies as well that for $T'_2$ = $-T_2$ we set $T_2$ = $T_1$.
 
Therefore the gluon polarization four-vector is $e$ = 
$2T_2$ $l$ for $T'_2$ = $T_2$ and $e$ = -$k$ $\delta_{T_1,T_2}$ for $T'_2$ = 
$-T_2$, where, in the reference frame defined in subsubsect. A.1.4, 
$l$ $\equiv$ $(0,{\bf i})$ and $k$ $\equiv$ $(0,{\bf k})$. 
\vskip 0.6in
\setcounter{equation}{0}
\renewcommand\theequation{C.\arabic{equation}}

\appendix{\large \bf Appendix C}

Here we write the hadronic tensor according to quantum field theory, up to 
first order in $g$, adopting the axial gauge $A^a \cdot n_2 = 0$, with $a$ = 1
.... 8:

\begin{equation}
H^{G,f}_{\mu\nu} = \frac{1}{3} \int\int d^4 x d^4 y e^{-iqx} 
[\Gamma^{2f}_{\sigma\sigma'} (y)
\tilde{\Gamma}^{1f}_{\alpha\beta\beta'} (x, y) + (1\leftrightarrow 2)]
[\gamma^{\alpha} S(x-y) 
\gamma_{\mu}]^{\beta\sigma}\gamma_{\nu}^{\beta'\sigma'},
\label{qft}
\end{equation}
where

\begin{eqnarray}
\tilde{\Gamma}^{lf}_{\alpha\beta\beta'} (x, y) &=& 
<P_l|{\overline \psi}^f_{\beta'i}(0)
D^{ij}_{\alpha}(y) \psi^f_{\beta j}(x)|P_l>, 
\label{sof1}
\\
D^{ij}_{\alpha}(y) &=& i \partial_{\alpha} \delta^{ij} + g A^a_{\alpha}(y) 
{1\over 2} \lambda_a^{ij},
\label{sof2}
\\
\Gamma^{lf}_{\sigma\sigma'} (y) &=& <P_{\overline l}|\psi^f_{\sigma
i}(0) {\overline \psi}^f_{\sigma'i}(y)|P_{\overline l}>, 
\label{sof3}
\\
S(x-y) &=& \int {{d^4q'}\over{(2\pi)^4}} {{i \rlap/q'}\over{q'^2+i \epsilon}}
e^{-iq'(y-x)}.
\end{eqnarray}
Two observations are in order. First of all in the above expression we have 
omitted the chronological products, according to the considerations by various 
authors\cite{dgl} (see also\cite{bo4}). Secondly, the covariant derivative in 
the expression (\ref{sof1}) implies gauge invariance of the tensor (\ref{qft}).
However we have omitted 
the gauge "link" operators\cite{bo3} in the expressions (\ref{sof1}) and 
(\ref{sof3}), since in the gauge adopted these operators can be made unity. 
All this simplifies a lot the calculations and immediately allows to 
establish symmetry properties of the correlation functions (see subsect. 
\ref{subsec:sym}).

We consider the usual Fourier expansions of the of the massless fermion and 
vector boson fields, {\it i.e.},

\begin{eqnarray}
\psi^f_{\beta i}(x) = {1 \over {(2\pi)^{3/2}}}\sum_{T=-1/2}^{1/2} 
\sum_{i=1}^3 \int 
{{d^3p}\over{2^{1/2}|{\bf p}|^{1/2}}} [c^f_{T,i} ({\bf p}) u^T_{\alpha} 
({\bf p}) \chi_i e^{ipx}\nonumber\\
+d^{f\dagger}_{T,i} ({\bf p}) v^T_{\alpha} ({\bf p}) 
{\overline \chi}_i e^{-ipx}],
\label{fi1}
\\
A^c_{\alpha}(y) = {1 \over {(2\pi)^{3/2}}} \sum_{m = 1}^{2} \int  
{{d^3k}\over{2^{1/2} |{\bf k}|^{1/2}}} [a^c_m ({\bf k}) e_{\alpha}^m 
({\bf k})  e^{iky}+c.c.].
\label{fi2}
\end{eqnarray}
Here $e^m$ ($m$ $=$ 1, 2)  are the two different polarization four-vectors of 
the physical gluons: $e^1$ $=$ $S$ ({\it i. e.}, along the proton spin)
and $e^2$ $=$ $\bar{S}$  ({\it i. e.}, orthogonal to the proton spin 
and to the proton momentum). $a^c_m$ and $a^{c\dagger}_m$ are the creation and 
destruction operators of linearly polarized gluons.

Substituting expressions (\ref{fi1}) and (\ref{fi2}) into eq. (\ref{qft}), we 
obtain a tensor of the type

\begin{equation}
H^{G,f}_{\mu\nu} = H^{G,f}_{0\mu\nu} + g (H^{G,f,a}_{1\mu\nu}
\ - H^{G,f,b}_{1\mu\nu}).
\end{equation}
Here, recalling the definition 

\begin{equation}
\sum_{i=1}^3<P_l|c^{f\dagger}_{T_l,i} ({\bf p}_l) c^f_{T_l,i} ({\bf p}_l) 
\ |P_l> = q^f_{l T_l}  (x_l, {\bf p}_{l\perp}^2),
\label{dens}
\end{equation}
and an analogous one for antiquark densities, we find $H^{G,f}_{0\mu\nu}$ = 
$H^f_{\mu\nu}$ (see eq. (\ref{ht2}) in the text). Moreover we can identify  
$H_{1\mu\nu}^{G,f,a(b)}$ with $H_{\mu\nu}^{'f,a(b)}$ (cfr. eq. (\ref{htt4})), 
provided we normalize the quark correlation functions according to eq. 
(\ref{cort}) and assume a similar formula for the antiquark correlation 
functions. We stress that this normalization guarantees {\it gauge invariance}.

\vspace {1.5 cm}

\end{document}